# Low Power Superconducting Microwave Applications and Microwave Microscopy

Steven M. Anlage, C. P. Vlahacos, D. E. Steinhauer, S. K. Dutta,
B. J. Feenstra, A. Thanawalla, and F. C. Wellstood

Physics Department, Center for Superconductivity Research,
University of Maryland, College Park, MD  20742-4111  USA
http://www.csr.umd.edu

**Abstract**

We briefly review some non-accelerator high-frequency applications of superconductors. These include the use of high-$T_c$ superconductors in front-end band-pass filters in cellular telephone base stations, the High Temperature Superconductor Space Experiment, and high-speed digital electronics.  We also present an overview of our work on a novel form of near-field scanning microscopy at microwave frequencies.  This form of microscopy can be used to investigate the microwave properties of metals and dielectrics on length scales as small as 1 µm.  With this microscope we have demonstrated quantitative imaging of sheet resistance and topography at microwave frequencies.  An examination of the local microwave response of the surface of a heat-treated bulk Nb sample is also presented.

## I. Introduction

The discovery of high temperature superconductivity in 1986 created the hope of finding many new applications for superconductors in the everyday world.[1]  In this paper we discuss several applications which make use of the unique high frequency properties of superconductors.  We also present an overview of our recent work on near-field microwave microscopy, and how these instruments can have an impact on superconducting microwave technology.

## II.  High Frequency Applications of Superconductors

### II. A.  Cellular Telephone Front End Filters

One promising application for high temperature superconductors (HTS) is in filters for cellular telephone base stations.[2,3]  These stations must communicate with low-power transmitters within their assigned cell.  There are competing concerns which govern the size of a cell.  Fewer base stations are clearly more economical, but signal strengths at the base station are smaller, and the probability of dropped calls increases.  If the base station receiver system can be made more sensitive, or if the front-end filtering can have a reduced noise figure, the cell area can be made larger.  Another concern is the abundance of interfering signals present in the cellular communication frequency bands (800 – 900 MHz, and just below 2 GHz).  Interfering signals can come from neighboring users in the spectrum, or from multiple users in a given frequency band.

Filters made from HTS materials can help to solve both of the above problems.  Band pass filters are made by patterning coupled multiple pole resonators in thin films of HTS on high dielectric constant substrates.  The structures are very compact and have very little insertion loss when operated at temperatures less than about two-thirds of the superconducting transition temperature, $T_c$.  Because the filter must be cooled, one has the opportunity to add other cryogenic devices to the front end, such as a cooled low-noise semiconductor amplifier.  The combination of reduced insertion loss and lower noise figure in the amplifier make superconducting front-end filters very attractive for use in cellular networks.

Another attractive aspect of HTS front-end filters is their reduced size compared to conventional copper filters. Multiple pole filters can be made from HTS thin films on a 2-inch diameter LaAlO$_3$ substrate with a dielectric constant $\varepsilon_r \cong 23$. Even with the cryocooler, the HTS filters are significantly smaller than conventional filters. This size reduction is especially important in urban base stations where space comes at a premium.

Although sales of HTS cellular filters have been steady, there is some reluctance to adopt these systems for wide spread commercial use. One factor limiting their acceptance is the lack of long-term data on the performance of cryocoolers. The cryocoolers must work reliably and maintain a fixed temperature for at least 5 years with no regular maintenance. In addition, the cryogenic packaging must maintain its integrity over that time scale so that there is no interruption of service due to a vacuum leak and subsequent cryogenic failure.

Another limitation of current HTS filters is their high-power performance. Since present HTS thin films are granular, they show a significant increase in loss and inductance at high power. This power dependence is often blamed on the creation and motion of Josephson vortices at weak links in the material.[4] However, no direct experimental evidence exists on the microscopic origins of nonlinearity, so the question remains open. Improved deposition techniques have produced HTS films with dramatically better power-handling capability.[5,6] However, the thin film geometries required to achieve a given electrical performance often introduce regions in the device where very large current densities must flow. For example currents flowing through a flat strip must have very large current densities at the edges to screen out the self-induced magnetic field. These areas of the device are quite susceptible to rf breakdown. New filter designs, which do not involve the flow of current along patterned edges, are being developed to overcome these geometrical limitations.[7,8]

**II. B. The High Temperature Superconductor Space Experiment (HTSSE)**

The use of superconductors for high frequency communications is not limited to terrestrial applications. Significant advantages could be gained by using HTS microwave filters and other components on communications satellites. In this application, weight is a major issue. If superconducting components with equivalent performance can be produced which weigh less and are as reliable as their normal metal counterparts, they will be favored for satellite communications applications.

However, satellite manufacturers are conservative and do not wish to jeopardize an entire communications satellite to save weight with an unproven new technology. To address this issue, the Naval Research Laboratory started the HTSSE program.[9] The idea is to build HTS microwave components and subsystems which perform the same tasks as those required on a communications satellite, but to launch them on a dedicated test platform.

The first satellite HTSSE-I was assembled from superconducting components manufactured by several industrial research laboratories and one University.[9] These components included multiple-pole filters, delay lines, high-Q resonators, a stabilized superconducting low phase-noise oscillator, switchable and tunable filters, and a 3 dB coupler. The satellite was also designed to test the reliability and operation of the cryocooler which cooled the devices to 77 K. The satellite incorporated a space-qualified scalar network analyzer, which was designed to test each component in turn during the flight and send the data back to the ground for further analysis. The experiment was designed to test the ability of HTS devices to withstand the rigors of reaching orbit and operating there, as well as to quantify the improved performance in a hostile environment. Unfortunately the payload did not reach orbit and the experiment was lost.

While the HTSSE-I project was focused on simple HTS devices, the HTSSE-II program is designed to address more complex HTS components and sub-systems.[10] The objective is to maintain the experiment in orbit for 1 year at a temperature of 77 K. The configuration of this satellite is to test a full superconducting "bent pipe" communications system. The signal is received by a normal metal antenna and sent to a primarily superconducting front-end system. The system includes a 4 channel input multiplexer and 4 channel filter, both operating at 4 GHz, two receivers (one superconducting and the other a super/semiconductor hybrid), an analog-to-digital converter using HTS Josephson junctions, a 5-bit HTS digital frequency measurement system, a digital multiplexer, and a superconducting delay line. The satellite has been assembled and is now awaiting launch. To demonstrate a more complete HTS communications system, a third satellite may be built in the future.

## II. C. Rapid Single Flux Quantum Superconducting Digital Electronics

An emerging high frequency application of superconductors comes from an unlikely direction: digital electronics. Semiconductor-based digital computers are moving to higher clock frequencies at a relentless pace, but are still operating below 1 GHz. By contrast, superconducting digital electronics has pushed clock frequencies to hundreds of GHz through the use of Josephson effects. A single Josephson junction can undergo a $2\pi$ phase shift on time scale less than 1 ps. This fundamentally fast process has been used as the heart of superconductive digital electronics for several decades. However, early designs made use of latching logic, in which a Josephson junction was sent from the zero resistance to a finite voltage state upon receiving a voltage pulse. This leads to several problems, including relatively large amounts of dissipated energy, the need to continually refresh all of the logic gates, and the relatively slow process of making a Josephson junction turn off and go back into the zero resistance state.[11,12]

To overcome these problems, rapid single flux quantum (RSFQ) superconducting logic was developed. This form of digital logic still uses the Josephson junction as the basic circuit element, but takes full advantage of the unique high frequency properties of superconductors. The first idea is to use the voltage pulse from a moving single flux quantum as the unit of information. The short voltage pulse V(t) corresponds to a single flux quantum moving across a Josephson junction (i.e. a $2\pi$ phase flip). It has the property that $\int V(t)dt = \Phi_0 = 2.07$ mV-ps, the quantum of magnetic flux. The pulses typically have a maximum amplitude on the order of the gap voltage ($2\Delta/e \sim 1$ mV for Nb) and a duration of a few ps. The second important idea is to use superconducting transmission lines to move the RSFQ pulses from one logic element to another. Superconducting transmission lines have low dispersion and low loss, thus preserving the spectral content of the short pulses.[13,14]

Using these two basic ingredients, RSFQ circuits can be designed to reproduce, amplify, and store voltage pulses. More complicated circuits, such as flip-flops, can also be made; an RSFQ flip-flop operating at 350 GHz has been demonstrated. RSFQ circuits are quite useful in performing fast analog-to-digital conversion.[15] Workers have achieved 3-bit A/D conversion at a frequency of 20 GHz, and 4-bit D/A at 1 GHz.[16] The circuits have been implemented in both low-$T_c$ and high-$T_c$ materials and continue to be a focus of active research and development.[15]

## III. Near-Field Scanning Microwave Microscopy

The combination of scanned probe microscopy and near-field electromagnetic sources has produced a powerful new set of research instruments for physics and materials science. Although much attention has been paid to near-field scanning optical microscopes (NSOM),[17] we have developed a class of near field microscopes which operate at rf, microwave, and millimeter-wave frequencies.[18] Our microscopes explore the physics, materials properties, and device characteristics in a technologically important part of the electromagnetic spectrum between 100 MHz and 50 GHz. We have demonstrated imaging with spatial resolution more than one thousand times smaller than the free-space wavelength of the radiation,[19] and we have developed

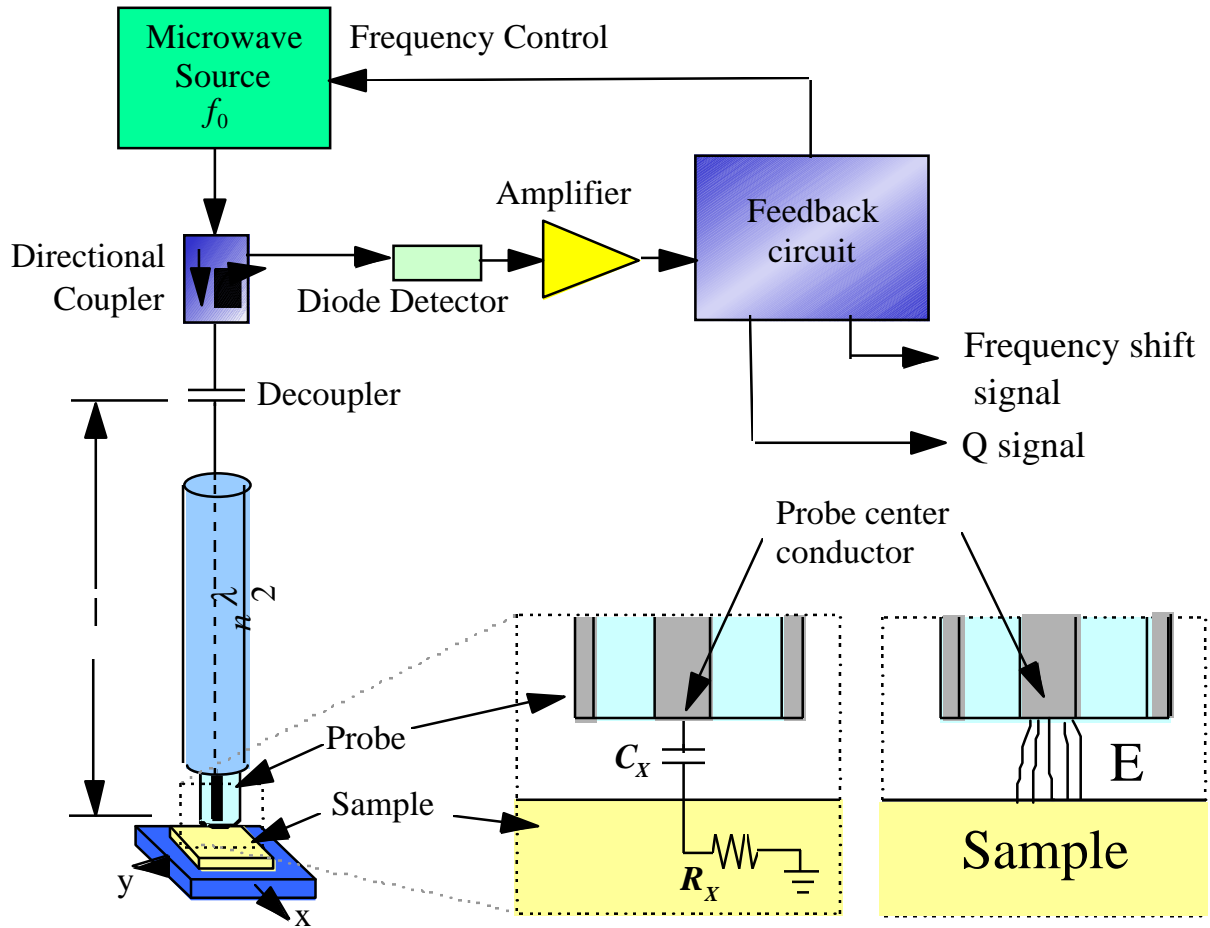

Fig. 1 Schematic of our scanning near-field microwave microscope. The microwave source is frequency locked to one of the resonant modes of the coaxial transmission line resonator. The sample perturbs one end of the resonator, and the resulting change in resonant frequency and quality factor are recorded as a function of the position of the probe over the sample.

quantitative measurements of surface sheet resistance [20,21] and topography at microwave frequencies.[22]

Near field microwave measurements have been pursued by many groups over the years. The earliest work by Soohoo [23] and Bryant and Gunn [24] used scanned resonators with small apertures to couple to the sample of interest. The method of Ash and Nichols used an open resonator formed between a hemispherical and planar mirror.[25] They opened a small hole in the planar mirror allowing an evanescent wave to leave the cavity (like Soohoo), and scanned a sample beneath it. As this resembles the configuration now used for NSOM, it is often reported to be the first modern near-field evanescent electromagnetic wave microscope. Near-field imaging has been accomplished using evanescent waves from the optical to the microwave range in coaxial, waveguide, microstrip, and scanning tunneling microscope geometries [26-36].

### III. A. Near-Field Microwave Materials Evaluation

Over the past two years, our group has developed a novel form of near-field scanning microwave microscopy. As shown in Fig. 1, our microscope consists of a resonant coaxial cable which is weakly coupled to a microwave generator on one end through a decoupling capacitor $C_d$, and coupled to a sample through an open-ended coaxial probe on the other end. As the sample is scanned beneath the probe, the probe-sample separation will vary (depending upon the topography of the sample), causing the capacitive coupling to the sample, $C_x$, to vary. This has the result of changing the resonant frequency of the coaxial cable resonator. Also, as the local sheet resistance ($R_x$) of the sample varies, so will the quality factor, Q, of the resonant cable. A circuit is used to force the microwave generator to follow a single resonant frequency of the cable, and a second circuit is used to measure the Q of the circuit, both in real time.[20,21] Hence as the sample is scanned below the open-ended coaxial probe, the frequency shift and Q signals are collected, and corresponding two-dimensional images of the sample topography and sheet resistance can be generated. The circuit runs fast enough to accurately record at scan speeds of up to 25 mm/sec.

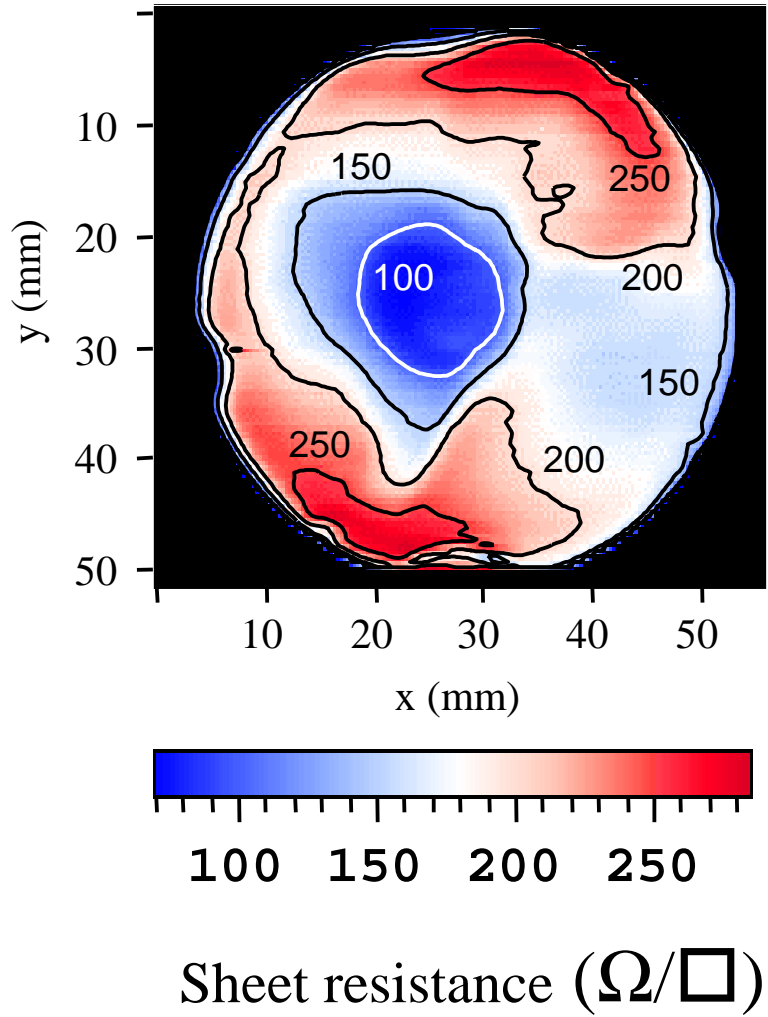

Sheet resistance ($\Omega/\square$)

Fig. 2  Calibrated sheet resistance image of an $YBa_2Cu_3O_7$ thin film deposited on a 2-inch diameter sapphire wafer. The image was acquired at 7.5 GHz with a 480 µm diameter probe at a height of 50 µm at room temperature.

The spatial resolution of such a microscope has been demonstrated to be the larger of the probe-sample separation and the diameter of the inner conductor wire in the open-ended coaxial cable.[19] As with other forms of near-field microscopy, the probe is placed well within one wavelength of the sample under study. This is particularly easy to accomplish at rf, microwave, and millimeter-wave frequencies because the wavelength ranges from meters to millimeters. However, the spatial resolution of our microscopes also depends on the probe-sample separation. To achieve fine resolution, we must place the probe closer to the sample, and maintain a constant separation. Using our current instruments, we have achieved a spatial resolution of approximately 100 µm without contacting the sample, and less than 2 µm in contact mode.

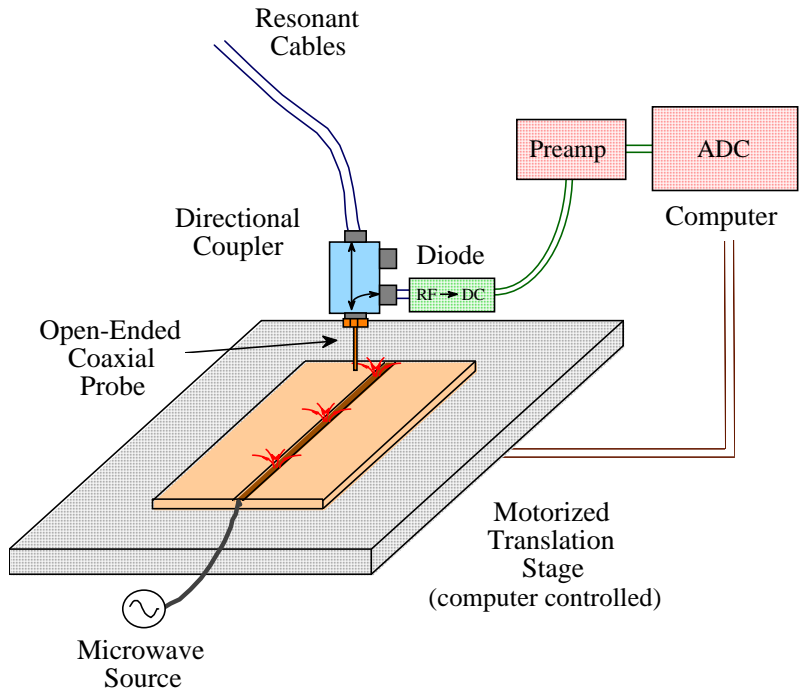

Fig. 3 Schematic illustration of scanning near-field microwave microscope operated as an electric field microscope over an active device. The open-ended coaxial probe is sensitive of the electric field normal to the surface of the inner conductor.

We have demonstrated that our scanning near-field microwave microscope can be used to obtain quantitative topographic images with 55 nm vertical sensitivity,[22] and quantitative surface sheet resistance images of metallic thin films.[21] Figure 2 shows a sheet resistance image of a $YBa_2Cu_3O_7$ thin film deposited on a 2-inch diameter sapphire wafer. The film shows a sheet resistance which varies by a factor of 3 over its surface. The image was obtained at room temperature in only 10 minutes. The sensitivity of the microscope can be dramatically increased by using a superconducting coaxial cable.

We are presently developing microscopes to make quantitative dielectric constant measurements (both $\varepsilon_r$ and $\tan\delta$) at microwave frequencies, as well as complex permeability measurements of magnetic films, both at room temperature.

### III. B. Electromagnetic Field Imaging

The microscope can also be used to image electric and magnetic fields in the vicinity of operating microwave devices. Figure 3 illustrates how we use the open-ended coaxial probe to pick-up electric fields above an operating device. The signals are enhanced by the resonant Q of the system and recorded on a computer.[19]

We have imaged a variety of devices using this system. For example, Figure 4(a) shows a short microstrip transmission line lithographically defined on a two-sided copper printed circuit board which terminates at an open circuit on the right hand side. The strip is approximately 1 mm wide, and the dielectric is 0.5 mm thick. Figure 4(b) shows three active mode images of the microstrip taken at 8.03, 9.66 and 11.29 GHz with a signal applied through the coaxial cable on the left. The image is taken over the right two-thirds of the microstrip shown in Fig. 4(a). In each case there is a maximum signal on the right hand side at the position of the open termination demonstrating that, as expected, the microscope is sensitive to voltage rather than current. A clear standing wave pattern is seen in each image, with a wavelength which decreases linearly with increasing frequency. Analysis of the three images gives effective dielectric constants for the printed circuit board microstrip in the range of 3.32 to 3.46.

The microscope is sensitive to the component of electric field normal to the surface of the exposed inner conductor, in this case the component of electric field normal to the circuit plane. The image demonstrates clearly that there are standing waves on the transmission line, and will be of great assistance in the design of new planar microwave devices (both normal and superconducting). In addition, by performing measurements at various heights above the sample, we can reconstruct the field profile, and then calculate the electric potential.

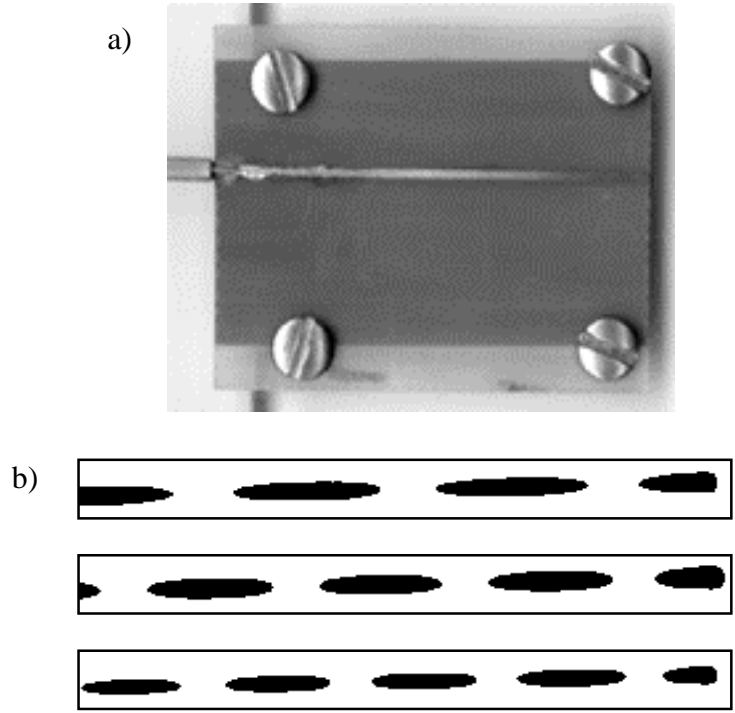

Fig. 4 a) Optical photograph of a copper microstrip circuit showing coax-to-microstrip transition on left, and open circuit termination on right. The printed circuit board is 39 mm long and 33 mm wide, and the line is approximately 1 mm wide. b) The lower part of the figure shows active images of the copper microstrip with standing waves present. Operating frequencies are (from top to bottom) 8.03 GHz, 9.66 GHz, and 11.29 GHz. The horizontal scale is 32.5 mm and the vertical scale on each panel is 3 mm.

### III. C. Microwave Imaging of Bulk Nb Surfaces

We have used a near-field scanning microwave microscope to image the high frequency properties of a bulk Nb surface. A Nb heat treatment was performed by Peter Kneisel of the Thomas Jefferson National Accelerator Laboratory,[37] and the sample was subsequently etched several times with an aggressive $HF/HNO_3$ acid treatment. Optical micrographs of the surface show irregular grains with sizes ranging from 400 µm to about 1.5 mm in diameter, with most on the order of 1 mm in diameter. Upon closer examination, inside the grains one finds surface structures which are approximately 10 µm across. On some grains these structures are pyramidal in shape, while in others they are dome-like, resembling bubbles. The grain boundaries were also deeply etched and have a total width of approximately 10 µm on the surface.

We imaged the bulk Nb surfaces with a 480 µm inner conductor diameter (see Fig. 5). On the left side of Fig. 5 is an image of the 2f signal, which is proportional to the microscope quality factor, Q.[21] The image on the right side of Fig. 5 shows the frequency shift of the microscope, and gives information about the surface resistance[20] and surface topography.[22] The image shows a cleared hole in the Nb piece, as well as a scratch on the surface approximately 3 mm long and several hundred microns deep. Both features are easily imaged and produce excellent contrast in Fig. 5. Next we investigated a "featureless" 2 x 2 mm area of the Nb surface. The near-field microwave microscope images shown in Fig. 6 were obtained with a 200 μm inner conductor diameter probe operating at a height of approximately 5 μm and a frequency of 7.47 GHz. Fig. 6 shows that the 2f (proportional to resonator Q) and frequency shift images are quite similar, showing common features which are on the scale of the probe inner conductor diameter, or greater. The 2f image shows regions of low Q on length scales less than a typical grain size, which may be associated with localized areas having a lateral extent less than 200 μm. (Note that the spacing between tick marks in Fig. 6 shows the approximate resolution of the probe.) The 2f image also shows larger areas, on the order of a typical grain size, with lower Q. The frequency shift image reproduces many of the features seen in the 2f image.

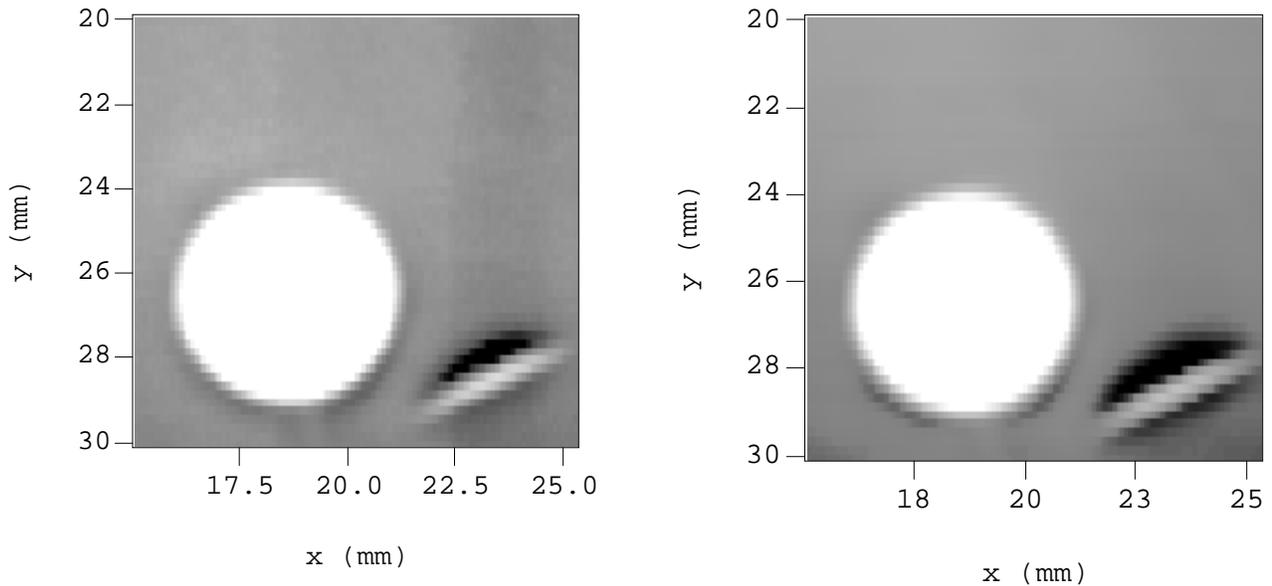

Fig. 5 Q (left) and frequency shift (right) microwave microscope images of a cleared hole and 3 mm long scratch in the surface of a bulk Nb sample. The images were obtained with a 480 μm inner conductor diameter probe at a frequency of 7.46 GHz and a height of approximately 50 μm, and cover a 10 mm x 10 mm area. The dark areas correspond to lower Q and more negative frequency shift in the left and right images, respectively.

It should be noted that the dark regions in Fig. 6 show both a decrease in Q and a downward frequency shift. There are two scenarios for the surface features which produce this kind of image. The first scenario consistent with this behavior is that the surface has a uniform surface resistance and variable surface topography. The dark areas in Fig. 6 would be interpreted as raised features on the surface. A smaller probe / sample separation produces a downward shift in resonant frequency and Q, independent of surface resistance.[20] In this interpretation, Fig. 6 shows simply the topography of a uniformly conducting surface.

The second scenario is that the surface has a flat topography with variable surface resistance. The dark features in Fig. 6 would then correspond to regions of lower surface resistance, since frequency shift monotonically becomes more negative as $R_s$ decreases at fixed height. To be consistent with the known Q vs. $R_s$ curve for this probe,[20,21] the surface sheet resistance must exceed approximately 400 Ω/sq. Only in this range of sheet resistance, does the Q decrease with decreasing sheet resistance.[20] In this interpretation most of the Nb surface has a sheet resistance considerably greater than 400 Ω/sq, with pockets of somewhat lower sheet resistance.

Since the sheet resistance of bulk Nb is expected to be on the order of $10^{-3}$ Ω at this frequency, the topographic interpretation of the images is almost certainly the correct one. More detailed examinations of heat-treated surfaces are planned for the future.

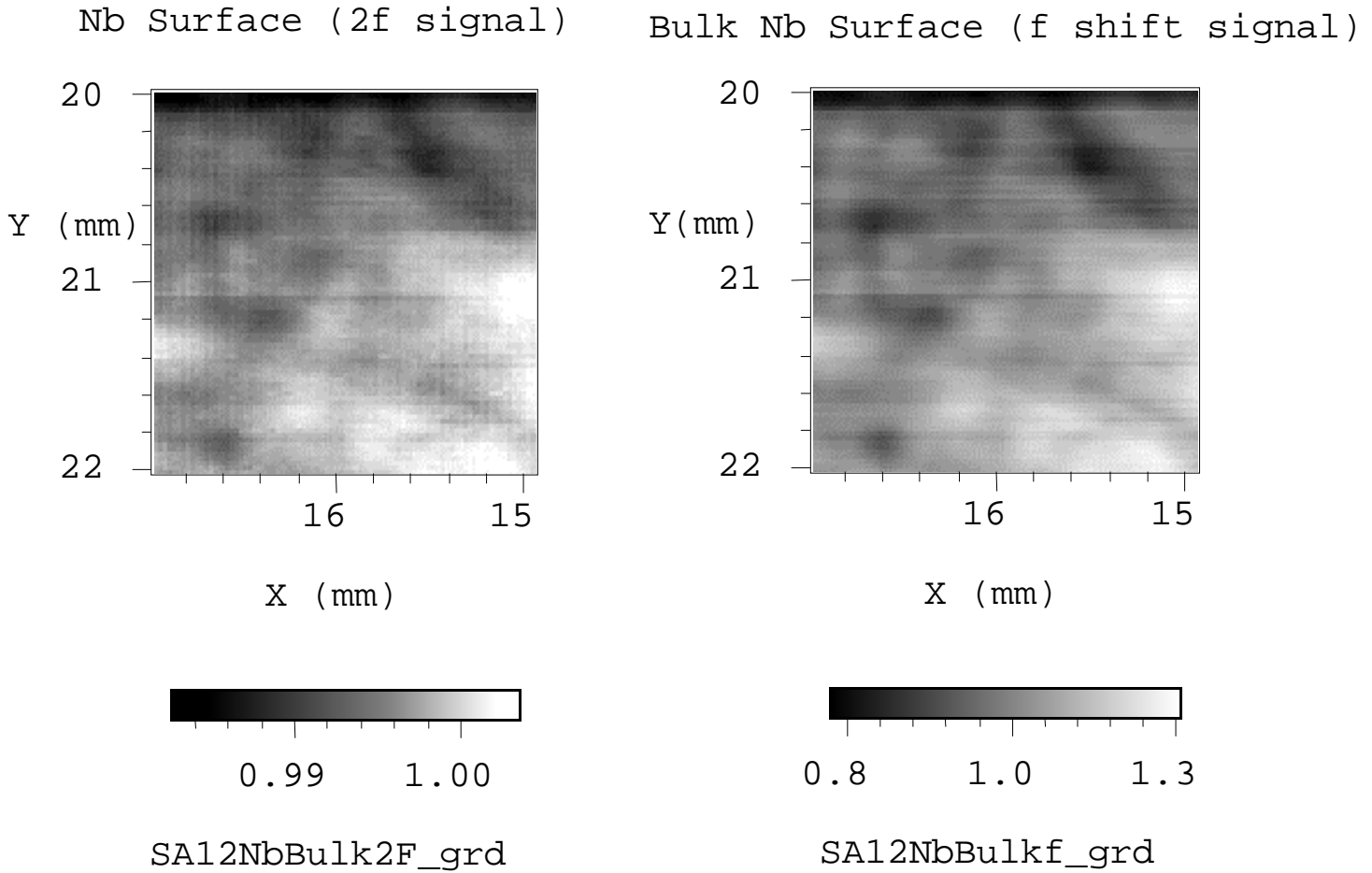

Fig. 6 Microwave microscope images of the surface of bulk Nb. These scans cover an area of 2 x 2 mm and show the 2f signal (proportional to Q of the microscope) on the left, and the frequency shift on the right.

## IV. Conclusions

It is now widely believed that high frequency applications of HTS superconductors offer the most potential for commercialization and common everyday use. Much progress has been made on the science and technology of HTS superconductor high frequency applications. The near-field scanning microwave microscope will be a useful tool for characterizing the microwave properties of these materials, and for evaluating the distribution of electromagnetic fields around superconducting devices.

## V. Acknowledgements

We have benefited from conversations with Peter Kneisel, Michael Pambianchi, and Marty Nisenoff. This work is supported by the Maryland/NSF Materials Research Science and Engineering Center on Oxide Thin Films (NSF DMR-9632521), NSF DMR-9624021, NSF ECS-9632811, and the Maryland Center for Superconductivity Research.

## VI. References

[1] G. B. Lubkin, Physics Today 48 (3) 20 (1995).
[2] M. J. Scharen, D. R. Chase, A. M. Ho, A. O'Baid, K. R. Raihn, and R. J. Forse, IEEE Trans. Appl. Supercond. 7, 3744 (1997).
[3] R. Hammond, Hey-Shipton, and G. Matthaei, IEEE Spectrum, April (1993).

[4]  J. Halbritter, J. Superconductivity $\underline{8}$, 691 (1995).
[5]  D. W. Face, C. Wilker, Z.-Y. Shen, P. Pang, and R. J. Small, IEEE Trans. Appl. Supercond. $\underline{5}$, 1581 (1995).
[6]  D. W. Face, *et al*., IEEE Trans. Appl. Supercond. $\underline{7}$, 1283 (1997).
[7]  Z.-Y. Shen, C. Wilker, P. Pang, D. W. Face, C. F. Carter, and C. M. Harrington, IEEE Trans. Appl. Supercon. $\underline{7}$, 2446 (1997).
[8]  G. Muller, *et al*., , IEEE Trans. Appl. Supercon. $\underline{7}$, 1287 (1997).
[9]  See the special issue of IEEE Trans. Microwave Theory Tech. $\underline{39}$ (9) (1991).
[10]  See the special issue of IEEE Trans. Microwave Theory Tech. $\underline{44}$ (7) (1996).
[11]  K. K. Likharev and V. K. Semenov, IEEE Trans. Appl. Supercon. $\underline{1}$, 3 (1991).
[12]  K. K. Likharev, in *The New Superconducting Electronics*, edited by H. Weinstock and R. W. Ralston (Kluwer, Amsterdam, 1993) p. 423.
[13]  J. C. Swihart, J. Appl. Phys. $\underline{32}$, 461 (1961).
[14]  R. E. Matick, *Transmission Lines for Digital and Communication Networks*, (McGraw-Hill, New York, 1969), p. 211.
[15]  See any recent proceedings of the Applied Superconductivity Conference, such as IEEE Trans. Appl. Supercon. $\underline{7}$ (2) (1997).
[16]  R. D. Sandell, B. J. Dalrymple, and A. D. Smith, IEEE Trans. Appl. Supercon. $\underline{7}$, 2468 (1997).
[17]  E. Betzig and J. K. Trautman, Science $\underline{257}$, 189 (1992).
[18]  C. P. Vlahacos, R. C. Black, S. M. Anlage, and F. C. Wellstood, Appl. Phys. Lett. $\underline{69}$, 3272 (1996).
[19].  Steven M. Anlage, C. P. Vlahacos, Sudeep Dutta, and F. C. Wellstood,  IEEE Trans. Appl. Supercond. $\underline{7}$, 3686 (1997).
[20]  D. E. Steinhauer, C. P. Vlahacos, Sudeep Dutta, F. C. Wellstood, and Steven M. Anlage, Appl.   Phys. Letts. $\underline{71}$, 1736 (1997).
[21]  D. E. Steinhauer, C. P. Vlahacos, S. K. Dutta, B. J. Feenstra, F. C. Wellstood, and Steven M. Anlage, to be published in Applied Physics Letters, early 1998.
[22]  C. P. Vlahacos, D. E. Steinhauer, S. K. Dutta, B. J. Feenstra, Steven M. Anlage, and F. C. Wellstood, Submitted to Appl. Phys. Lett., November 25, 1997.
[23]  R. F. Soohoo, J. Appl. Phys. $\underline{33}$, 1276 (1962).
[24]  C. A. Bryant and J. B. Gunn, Rev. Sci. Instrum. $\underline{36}$, 1614 (1965).
[25]  E. A. Ash and G. Nichols, Nature $\underline{237}$, 510 (1972).
[26]  U. Durig, D. W. Pohl and F. Rohmer, J. Appl. Phys. $\underline{59}$, 3318 (1986).
[27]  R. J. Gutman, J. M. Borrego, P. Chakrabarti and Ming-Shan Wang, IEEE MTT-S Digest, pp 281 (1987).
[28]  M. Tabib-Azar, N. S. Shoemaker and S. Harris, Meas. Sci. Tech. $\underline{4}$, 583 (1993).
[29]  T. Wei, X. D. Xiang, W. G. Wallace-Freedman, and P. G. Schultz, Appl. Phys. Lett. $\underline{68}$, 3506 (1996).
[30]  M. Golosovsky, A. Galkin, and D. Davidov, IEEE Trans. Microwave Theory Tech. $\underline{44}$, 1390 (1996).
[31]  G. Nunes, and M. R. Freeman, Science $\underline{262}$, 1029 (1993).
[32]  R. J. Hamers, and D. G. Cahill, Appl. Phys. Lett. $\underline{57}$, 2031 (1990).
[33]  W. Seifert, E. Gerner, M. Stachel, and K. Dransfeld, Ultramicroscopy $\underline{42-44}$, 379 (1992).
[34]  S. J. Stranick, L. A. Blumm, M. M. Kamma, and P. S. Weiss, in Photons and Local Probes, edited by O. Marti and R. Mšller (Kluwer, Netherlands, 1995), p. 221.
[35]  F. Keilmann, D. W. van der Weide, T. Eickelkamp, R. Merz, and D. Stakle, Opt. Commun. $\underline{129}$, 15 (1996).
[36]  M. Fee, S. Chu, and T. W. Hansch, Optics Commun. $\underline{69}$, 219 (1989).

[37] The heat treatment consisted of a 4 hour anneal at 1400 C in a furnace with a background pressure of 5 x $10^{-9}$ Torr.